\documentclass[pre,aps,preprint,showpacs]{revtex4-1}
\usepackage{revsymb4-1,latexsym,amssymb,amsmath,comment}
\usepackage{graphicx,psfrag,subfigure,stmaryrd}
\usepackage[ansinew]{inputenc}

\newcommand{\be}{\begin{equation}}
\newcommand{\ee}{\end{equation}}
\newcommand{\bel}[1]{\begin{equation}\label{#1}}
\newcommand{\bea}{\begin{eqnarray}}
\newcommand{\eea}{\end{eqnarray}}
\newcommand{\ba}{\begin{array}}
\newcommand{\ea}{\end{array}}
\newcommand{\bef}{\begin{figure}}
\newcommand{\ef}{\end{figure}}

\begin{document}

\author{Thomas Bose and Steffen Trimper}
\affiliation{Institute of Physics,
Martin-Luther-University, D-06099 Halle, Germany}
\email{thomas.bose@physik.uni-halle.de}
\email{steffen.trimper@physik.uni-halle.de}
\title{Lagrangian approach and dissipative magnetic systems}
\date{\today }

\begin{abstract}
A Lagrangian is introduced which includes the coupling between magnetic moments $\mathbf{m}$
and the degrees of freedom $\boldsymbol{\sigma}$ of a reservoir.  In case the system-reservoir coupling breaks
the time reversal symmetry the magnetic moments perform a damped precession around an effective field
which is self-organized by the mutual interaction of the moments. The resulting evolution equation
has the form of the Landau-Lifshitz-Gilbert equation. In case the bath variables are constant vector
fields the moments $\mathbf{m}$ fulfill the reversible Landau-Lifshitz equation. Applying Noether's theorem
we find conserved quantities under rotation in space and within the configuration space of the moments.
\end{abstract}

\maketitle


\section{Introduction}

\noindent The dynamics of magnetic systems is described in a wide range of time and length scales from
a quantum approach up to a macroscopic thermodynamic access. On a coarse-grained mesoscopic level the
relevant electronic degrees of freedom are grouped into effective magnetic moments. As the consequence
the magnetization is characterized by a spatiotemporal vector field $\mathbf{m}(\mathbf{r},t)$. Owing to
the mutual interaction between the magnetic moments they perform a precession motion around a local effective
field giving rise to the propagation of spin-wave excitations. Due to a system-reservoir coupling the precession
of the moments should be a damped one. To analyze this situation one has to specify the coupling between
the system and the bath. The most popular approach to incorporate dissipation is the embedding of the relevant
system into a quantum-statistical environment which is assumed to remain in thermal equilibrium. The reservoir
is often represented by harmonic oscillators or spin moments which are analyzed by path integral techniques
\cite{feynman:2010:Book:QMandPI,kleinert:2009:book:pathintegrals,weiss:1999:book:quantdissip}.
A specification of the path integral approach for spin systems can be found in \cite{Schulman:PhysRev176:1558:1968,
Cabra:JPA30:2699:1997, Smirnov:JPA32:1285:1999, Grinberg:PLA311:133:2003}. A more generic description
concerning dissipative semiclassical dynamics is presented in \cite{Koch:PhysRevLett.100.230402:2008}. Although
the application of path integrals can be considered as an intuitive formalism analytical calculations are
often impossible and numerical schemes are necessary.
\noindent In the present paper we propose an alternative way to include dissipative effects for mesoscopic
magnetic systems. On this level
the analysis of magnetodynamics is performed properly by applying
the Landau-Lifshitz-Gilbert equation designated as LLG \cite{Landau:ZdS:8:p153:1935,
Gilbert:ITOM:40:p3443:2004}. A comprehensive survey of magnetization dynamics is given in
\cite{Tserkovnyak:RoMP:77:p1375:2005}. Our investigation can be grouped in the effort to understand dissipative
mechanisms in magnets. So, a non-linear dissipative model for magnetic systems was discussed in
\cite{Durand:EPL89:67004:2010}. On the relation between fluctuation-dissipation theorems and damping terms
like that one occurring in the LLG was reported in \cite{Safonov:PhysRevB.71.224402:2005}. The dynamical response
of ferromagnetic shape memory alloy actuators can be modeled by means of a dissipative Euler-Lagrange equation as
performed in \cite{Weetman:JAP105:2009}. Likewise, the pinning of magnetic domain walls in multiferroics is
discussed in terms of the EL equations in \cite{Gareeva:EPL91:47006:2010}. An alternative ansatz is introduced in
\cite{Ovchinnikov:PhysRevB.82.024410}, where a Lagrangian density is obtained based on a projection onto
the complex plane. The procedure gives rise to a dynamical equation which is equivalent to the Landau-Lifshitz
equation.
\noindent Different to the mentioned approaches the present paper is aimed to derive an equation of motion for a
magnetic system on a mesoscopic scale under the influence of a bath which likewise consists of mesoscopic
moments. Following this idea we propose a Lagrangian comprising both fields, $\mathbf{m}(\mathbf{r},t)$ as the system
variables and $\boldsymbol{\sigma}$ as the bath variables. The bath becomes dynamically active by the coupling to the system.
In case the coupling between system and reservoir breaks the time reversal symmetry the motion of the moments
$\mathbf{m}(\mathbf{r},t)$ is damped. The Lagrangian is modified in such a manner that dissipation can occur.


\section{The Lagrangian}

\begin{figure}
\begin{center}
  \includegraphics[width=0.98\linewidth]{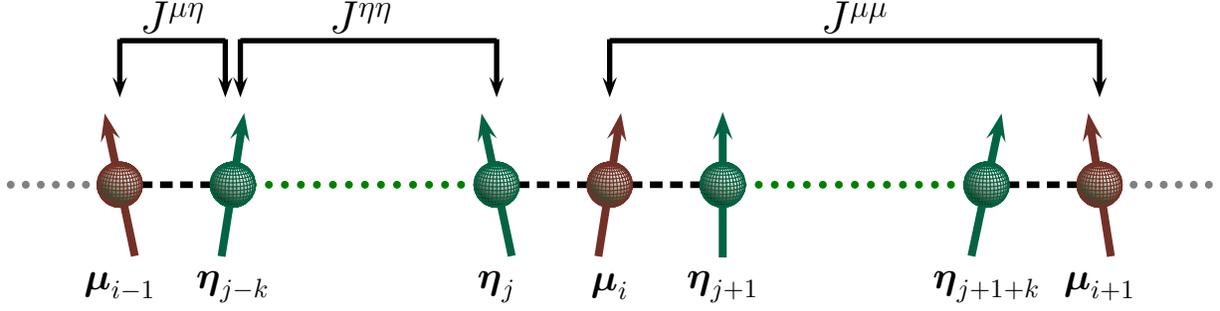}
  \caption{(Color online) Schematic illustration of the basic model. The red spins represent the magnetic moments $\mu_i$ and refer to the
lattice L1 introduced in the text. The green spin vectors $\eta_i$ build the bath lattice L2. Interactions are possible between the $\boldsymbol{\mu}_i$ and
$\boldsymbol{\eta}_j$, $\boldsymbol{\mu}_i$ and $\boldsymbol{\mu}_j$ and $\boldsymbol{\eta}_i$ and $\boldsymbol{\eta}_j$. The respective coupling strengths correspond to the
coupling parameters in Eqs.~\eqref{Lagm}-\eqref{Lagmsig} as follows: $J^{\mu \eta}\leftrightarrow {J}^{(m\sigma)}$, $J^{\mu \mu}\leftrightarrow {J}^{(m)}$ and
$J^{\eta \eta}\leftrightarrow {J}^{(\sigma )}$.}
  \label{BildSK}
 \end{center}
\end{figure}

\noindent As indicated we are interested to construct a Lagrangian describing the motion of a magnetization vector
field within a bath of spins. This reservoir should influence the measurable magnetization due to the mutual
interaction. Let us formulate the general assumptions for the underlying model. The magnetic order is originated by
single magnetic atoms which occupy equivalent crystal positions. Here we refer to a continuous description in
terms of a field vector denoted as $\mathbf{m}(\mathbf{r},t)$. Because the ferromagnet is considered below the Curie
temperature a sufficient number of microscopic spins preferring a parallel alignment are included in $\mathbf{m}$, i.e.
the effective magnetic moment is given by $\mathbf{m}(\mathbf{r},t)=\sum_i \boldsymbol{\mu}_i$
where the sum is extended over all microscopic moments within a small volume around the spatial coordinate $\mathbf{r}$ at time
$t$. As each axial vector the moment fulfills $\mathbf{m}(-t) = -\mathbf{m}(t)$. The bath in which the moments are embedded consists
likewise of mesoscopic spins. They are denoted as $\boldsymbol{\sigma}$ and are  also composed of microscopic moments
$\boldsymbol{\eta}_i$. This bath moments which play the role of 'virtual' moments are also axial vectors changing their sign by time inversion.
A further new aspect is that the coupling between the real and the virtual moments is not assumed to be weak. As the result
the complete system consists of two subsystems. One of them abbreviated as L1 is occupied exclusively by
the real spins with the moments $\mathbf{m}$ and the other one denoted as L2 is occupied by the bath spins $\boldsymbol{\sigma}$. The situation
is illustrated in FIG.~\ref{BildSK}.
Now let us introduce the action
\begin{align}\label{action}
    S[\{q_\alpha \}]= \int dt \int d^3x \,\mathcal{L}[\{q_\alpha \}] \,,
\end{align}
where the set $\{q_\alpha \}$ consists of the set of both moments $\boldsymbol{\sigma}$ and $\mathbf{m}$.
The Lagrange density comprises three terms
\begin{align}\label{Lag2}
\begin{aligned}
     \mathcal{L}[\mathbf{m} ,\dot{\mathbf{m}} , \nabla \mathbf{m};  \boldsymbol{\sigma},  \nabla \boldsymbol{\sigma}]  =  \mathcal{L}^{(m)}[\mathbf{m} , \dot{\mathbf{m}} ,
\nabla \mathbf{m}] + \mathcal{L}^{(\sigma)}[\nabla \boldsymbol{\sigma}] + \mathcal{L}^{(m\sigma)}[\boldsymbol{\sigma},\dot{\mathbf{m}}] \,,
\end{aligned}
\end{align}
where $\mathcal{L}^{(m)}$ indicates the Lagrangian of the magnetic system, $\mathcal{L}^{(\sigma)}$ represents the reservoir and the interaction term is denoted as
$\mathcal{L}^{(m\sigma)}$.
To be more specific the magnetic moments of the system interact via exchange coupling defined by the Lagrangian
\begin{align}\label{Lagm}
\mathcal{L}^{(m)}  = \frac{1}{2}{J}_{\alpha \beta }^{(m)} \frac{\partial m_\nu}{\partial x_\alpha }\frac{\partial m_\nu}{\partial x_\beta } + A_\nu (\mathbf{m} )\,\dot{m}_\nu\,,
\end{align}
where ${J}_{\alpha \beta }^{(m)}$ is the coupling parameter, diagonal in the isotropic case. The first term represents the energy density of the magnetic system.
Because we are not considering the acceleration of magnetic moments a term of the order $\dot{\mathbf{m}}^2$ is missing. Moreover, the magnetic moments perform a
precession around an effective magnetic field, which is self-organized by the mutual interaction. Therefore the vector potential $\mathbf{A}$ depends on the moments, i. e.
$\mathbf{A}= \mathbf{A}(\mathbf{m}(\mathbf{r},t))$. The coupling has the same form as the minimal coupling
in electrodynamics. The bath Lagrangian is defined in a similar manner as
\begin{align}\label{Lagsig}
\mathcal{L}^{(\sigma)} &= \frac{1}{2}{J}_{\alpha \beta }^{(\sigma)} \frac{\partial \sigma_\nu}{\partial x_\alpha }\frac{\partial \sigma_\nu}{\partial x_\beta } \,,
\end{align}
with the coupling constant ${J}_{\alpha \beta }^{(\sigma)}$. Eventually, the interacting part between system and bath is written as
\begin{align}\label{Lagmsig}
\mathcal{L}^{(m\sigma)} = {J}_{\alpha \beta }^{(m\sigma)} \frac{\partial m_\nu}{\partial x_\alpha }\frac{\partial \sigma_\nu}{\partial x_\beta } +
B_\nu (\boldsymbol{\sigma} )\,\dot{m}_\nu \,,
\end{align}
with the coupling strength ${J}_{\alpha \beta }^{(m\sigma)}$. The second term is constructed in the same manner as in Eq.~\eqref{Lagm}, where the potential
$\mathbf{B}(\boldsymbol{\sigma})$ will be specified below, see Eq.~\eqref{bnu}. The dynamics of the bath variable $\boldsymbol{\sigma} $ remains unspecified for the present, i.e. the Lagrangian does not include a term of the form $\propto \dot{\boldsymbol{\sigma}} $. Owing to the constraint, introduced in the next section,
the dynamically passive bath is sensitive to a change of the system variables $\mathbf{m}$ in such a manner that small variations of the system variables
$\mathbf{m}$ are related to small variations of $\boldsymbol{\sigma} $. This procedure leads to a coupling between bath and system so that the
time reversal symmetry is broken.


\section{Relation to the Landau-Lifshitz-Gilbert equation}

\noindent In this section we find the equation of motion for the magnetization $\mathbf{m}(\mathbf{r},t)$ from Eq.~\eqref{Lag2} combined with Eqs.~\eqref{Lagm}-\eqref{Lagmsig}.
Using the principle of least action it follows
\begin{align}\label{variation}
\begin{aligned}
\biggl[ \frac{\partial \mathcal{L}}{\partial \sigma_\beta} - \frac{\partial }{\partial x_\alpha}\frac{\partial \mathcal{L}}{\partial \left( \frac{\partial
\sigma_\beta}{\partial x_\alpha}\right) } \biggr] \delta \sigma_\beta 
    + \biggl[ \frac{\partial \mathcal{L}}{\partial m_\beta} - \frac{\partial }{\partial t}\frac{\partial \mathcal{L}}{\partial \dot{m}_\beta} -
\frac{\partial }{\partial x_\alpha}\frac{\partial \mathcal{L}}{\partial \left( \frac{\partial m_\beta}{\partial x_\alpha} \right) } \biggr] \delta m_\beta= 0 \,,
\end{aligned}
\end{align}
where $\delta m_\beta$ and $\delta \sigma_\beta$ are the small variations which drive the value for the action out of the stationary state. In general, one derives
a system of coupled partial differential equations. However, to proceed further let us impose a constraint on the system. A small variation of $\sigma_\beta$ should
be related to a small variation of $m_\beta$. Thus, we make the ansatz
\begin{align}\label{delta}
     \delta \sigma_\beta = - \kappa \, \delta m_\beta\,,\qquad \textrm{with}\qquad \kappa = \textrm{const} > 0\,.
\end{align}
Notice that this condition should be valid only locally but not globally. Insofar Eq.~\eqref{delta} is comparable
to an anholonom condition in mechanics.  Moreover relation \eqref{delta} is in accordance with the behavior of
the moments $\mathbf{m}$ and $\boldsymbol{\sigma}$ under time inversion. Physically the last relation means that
the bath reacts to a change of the system only temporarily. Because the system-reservoir coupling should
typically break the time reversal symmetry the expansion of the function $B_\nu (\boldsymbol{\sigma} )$ in
terms of $\sigma$ includes only odd terms. In lowest order we get from Eq.~\eqref{Lagmsig}
\begin{align}\label{bnu}
    B_\nu (\boldsymbol{\sigma})= - c\, \sigma_\nu\,,\qquad \textrm{with} \qquad c= \textrm{const}\,.
\end{align}
Due to Eqs.~\eqref{delta} and \eqref{bnu} the second term in
Eq.~\eqref{Lagmsig} is of the form $\propto \boldsymbol{\sigma}\cdot \dot {\mathbf{m}} $. Such a term is not
invariant under time reversal symmetry $t \to -t$.
As demonstrated below the broken time inversion invariance gives rise to damping effects.
Inserting Eqs.~\eqref{delta} and \eqref{bnu} into Eq.~\eqref{Lag2} and performing the variation according to
Eq.~\eqref{variation} we get
\begin{align}\label{performvar}
\begin{aligned}
    0= &\left( \frac{\partial A_\nu}{\partial m_\beta}-\frac{\partial A_\beta}{\partial m_\nu }\right) \dot{m}_\nu + c\, \dot{\sigma}_\beta + \kappa c \, \dot{m}_\beta \\  
     &- [\,J^{(m)}\, -  \kappa J^{(m\sigma)}\,]\, \nabla^2 m_\beta 
    - [\,J^{(m\sigma)}\,- \kappa J^{(\sigma)}\,]\, \nabla^2 \sigma_\beta\,.
\end{aligned}
\end{align}
Here we have assumed for simplicity that all coupling tensors ${J}$ are diagonal: \mbox{${J}_{\alpha \beta }=J \delta_{\alpha \beta }$}. The first term on the right
hand side in Eq.~\eqref{performvar} reminds of the field strength tensor in electrodynamics \cite{jackson:1999:book:classicalelectrodynamics}. Thus, we rewrite
\begin{align}\label{fieldtensor}
\begin{aligned}
    \left( \frac{\partial A_\nu }{\partial m_\beta }- \frac{\partial A_\beta }{\partial m_\nu }\right) \dot{m}_\nu \equiv F_{\beta \nu}\,\dot{m}_\nu = \Bigl[ \dot{\mathbf{m}} \times (\,\nabla_{\mathbf{m}} \times \mathbf{A}(\mathbf{m}))\Bigr]_\beta\,.
\end{aligned}
\end{align}
As mentioned above the vector function $\mathbf{A}(\mathbf{m})$ is regarded as vector potential which depends on space-time coordinates via the magnetic
moment $\mathbf{m}(\mathbf{r},t)$. In vector notation the last equation reads
\begin{align}\label{vecnotation}
\begin{aligned}
    \dot{\mathbf{m}} \times (\nabla_\mathbf{m} \times \mathbf{A}) = [\,J^{(m)} - \kappa J^{(m\sigma)}\,]\, \nabla^2 \mathbf{m}-\kappa c \, \dot{\mathbf{m}} - c \, \dot{\boldsymbol{\sigma}} +  [\,J^{m\sigma} - \kappa J^{(\sigma)}\,]\,
 \nabla^2 \boldsymbol{\sigma} \,.
\end{aligned}
\end{align}
If one is interested in weak excited states of a ferromagnet it is reasonable to assume that the direction of the magnetization in space changes slowly while its absolute value
is fixed, that is $\mathbf{m}^2=1$. Without loss of generality we have set the amplitude of $\mathbf{m}$ to unity. In order to proceed it is necessary to specify the
condition which should be fulfilled by the function $\mathbf{A}(\mathbf{m})$. Having in mind the LLG then we make the ansatz
$\nabla_\mathbf{m} \times \mathbf{A}(\mathbf{m})=g\, \mathbf{m}$, $g= \textrm{const}$. Based on these assumptions we get from Eq.~\eqref{vecnotation}
\begin{align}\label{Geq}
    \frac{\partial \mathbf{m}}{\partial t} = \frac{1}{g}\, \Biggl(\mathbf{m} \times \mathbf{H}_{\rm{eff}}\Biggr) - \frac{\kappa c}{g}\, \Biggl(\mathbf{m} \times \frac{\partial
\mathbf{m}}{\partial t}\Biggr)\,.
\end{align}
Here the effective field is given by the expression
\begin{align}\label{Heff}
\begin{aligned}
    \mathbf{H}_{\rm{eff}} = \Bigl( J^{(m)} - \kappa J^{(m\sigma)} \Bigr) \nabla^2 \mathbf{m} 
- c\, \frac{\partial \boldsymbol{\sigma}}{\partial t} + \Bigl( J^{(m\sigma)} - \kappa J^{(\sigma)} +  \Bigr) \nabla^2 \boldsymbol{\sigma}   \quad .
\end{aligned}
\end{align}
Eq.~\eqref{Geq} is nothing else than the Gilbert equation \cite{Gilbert:ITOM:40:p3443:2004} by relating the prefactors as follows
\begin{align}\label{prefactors}
    \gamma = -\frac{1}{g}\,,\qquad \alpha = -\frac{\kappa c}{g} = \kappa c \gamma\,,
\end{align}
where $\gamma$ and $\alpha$ are the gyromagnetic ratio and the Gilbert damping parameter, respectively. Since both parameters are positive quantities it follows that $g<0$
as well as $\kappa c>0$.
Further, Eq.~\eqref{Geq} can be converted into the form of the equivalent and widely used Landau-Lifshitz-Gilbert equation which reads
\begin{align}\label{LLG}
\begin{aligned}
    \frac{\partial \mathbf{m}}{\partial t} = -\frac{\gamma}{(1+\alpha)^2}\, (\mathbf{m} \times \mathbf{H}_{\rm{eff}})- \frac{\alpha \gamma}{(1+\alpha ^2)}\, \Bigl[\mathbf{m} \times (\mathbf{m} \times \mathbf{H}_{\rm{eff}})\Bigr]\,,
\end{aligned}
\end{align}
Both quantities $\gamma$ and $\alpha$ are still related to the model parameters by the expressions in Eq.~\eqref{prefactors} whereas the
the effective field $\mathbf{H}_{\rm{eff}}$ is given by Eq.~\eqref{Heff}. Now we want to analyze this expression and in particular,
to assign a physical meaning to the more or less ad hoc introduced quantity $\boldsymbol{\sigma}$. In doing so one can distinguish four different cases:\\
\noindent (i) The bath is not included which corresponds formally to $\boldsymbol{\sigma}$ is a constant vector depending neither on coordinates nor on time. Then
obviously all derivatives with respect to
the coordinates and the time of $\boldsymbol{\sigma}$ disappear in Eq.~\eqref{Heff} and consequently, the set of $\{q_\alpha \}$ in Eq.~\eqref{action} does not include
$\boldsymbol{\sigma}$. From here we conclude that the variation fulfills $\delta \boldsymbol{\sigma}=0$ in Eq.~\eqref{variation} which can be easily realized
setting $\kappa =0$, cf. Eq.~\eqref{delta}. Thus, the effective field in Eq.~\eqref{Heff} comprises the pure exchange interaction ${J}^{(m)}$ between the
magnetic moments and the damping term in Eq.~\eqref{Geq} is absent due to $\alpha =0$ in Eq.~\eqref{prefactors}. A constant bath field $\boldsymbol{\sigma}$
lead to the Landau-Lifshitz equation in the exchange interaction approach without damping, compare \cite{Landau:StatPhysPart2:Book:1980}. It describes the precession of
magnetic moments of an effective field which is self-organized by the mutual interaction of the moments.\\
(ii) $\boldsymbol{\sigma} = \boldsymbol{\sigma}(t)$ depends only on the time and not on the spatial coordinates. Regarding Eq.~\eqref{Heff} the effective field
is modified by two additional contributions, namely one proportional to $\nabla^2\mathbf{m}$, originated in the exchange interaction of the magnetic moments, and the
other one $\propto \dot{\boldsymbol{\sigma}}$. The latter one could be associated with an external time dependent field or, if $\boldsymbol{\sigma}$ points into a fixed
direction, gives rise to magnetic anisotropy. In that case the anisotropy axis is spatially constant but the amount of the anisotropy is changing in time. Such a situation
could be realized for instance when the ferromagnetic sample is excited by the irradiation with electromagnetic waves. As already mentioned the exchange coupling
${J}^{(m)}$ is supplemented by a term $-\kappa J^{(m\sigma)}$. In this manner the exchange interaction is influenced by the coupling between $\mathbf{m}$ and
$\boldsymbol{\sigma}$ although the spatial dependence of $\boldsymbol{\sigma}$ is not taken into account explicitly.\\
(iii) $\boldsymbol{\sigma} = \boldsymbol{\sigma}(\mathbf{r})$ depends only on the spatial coordinates and not on the time. In this case we first recognize that the
coupling strength ${J}^{(m)}$ in the term $\propto \nabla^2 \mathbf{m}$ is influenced in the same manner as in case when $\boldsymbol{\sigma} = \boldsymbol{\sigma}(t)$,
see the previous point. Different to the former cases the expression $\propto \nabla^2 \boldsymbol{\sigma}$ becomes important for the effective field
in Eq.~\eqref{Heff}. The appearance of this term suggests that spatial inhomogeneities of the surroundings of the magnetic system represented by $\mathbf{m}$
have to be incorporated into the effective field. It seems to be reasonable that the origin of this term is an inherent one and should not be led back to external
fields. As possible sources we have in mind local varying fields like inner and outer demagnetization fields as well as accessible fields created for instance by different
local temperatures.\\
(iv) $\boldsymbol{\sigma} = \boldsymbol{\sigma}(\mathbf{r},t)$ is the most general case. Then external as well as internal fields are captured in the model. Thus, the
effective field in Eq.~\eqref{Heff} can be rewritten as
    \begin{align}\label{Heff2}
        \mathbf{H}_{\rm{eff}}(\mathbf{r},t) = \mathbf{H}_{\rm{exch}}(\mathbf{r}) + \mathbf{h}(\mathbf{r},t)\,,
    \end{align}
where $\mathbf{H}_{\rm{eff}}$ consists of two parts. The term $\mathbf{H}_{\rm{exch}}=(J^{(m)}- \kappa J^{(m\sigma)})\nabla^2 \mathbf{m}$ is due to the
exchange interaction between the magnetic moments whereas $ \mathbf{h}(\mathbf{r},t)$ represents other possible influences as discussed under the points (ii) and (iii).
The function $ \mathbf{h}$ is related to the quantity $\boldsymbol{\sigma}$ by
\begin{align}\label{kleinh}
\begin{aligned}
    \mathbf{h}(\mathbf{r},t)  =  - c\, \frac{\partial \boldsymbol{\sigma}(\mathbf{r},t)}{\partial t} 
+ \left[ J^{(m\sigma)} -\kappa J^{(\sigma)} \right]\,
\nabla^2 \boldsymbol{\sigma}(\mathbf{r},t)\,.
\end{aligned}
\end{align}
Remark that the formerly introduced quantity $\boldsymbol{\sigma}$ is related to the physically relevant effective field by the first derivation
with respect to the time and the second derivation with respect to the spacial coordinates via Eq.~\eqref{kleinh}. This equation is an inhomogeneous diffusion equation
which can be generally solved by means of the expansion into Fourier series and the assumption of accurate initial and boundary conditions which depend on the actual
physical problem.


\section{Symmetry and Conservation}

\noindent After regarding the special example of the LLG we proceed with the investigation of more general aspects. The Lagrangian density allows to discuss the behavior under
space-time dependent group transformation. For this purpose we apply Noether's theorem \cite{Noether:Ges.Wiss.Goettingen:1918} to our model. To be more precise we
consider the conservation equation \cite{Hill:ConservationTheorems:1951}
\begin{align}\label{NT1}
\begin{aligned}
\frac{\partial}{\partial \mathcal{X}_\alpha }\biggl[ \Bigl( \mathcal{L}\delta _{\alpha \beta } - \frac{\partial \mathcal{L}}{\partial (\partial_\alpha  \Psi_\gamma  )}
\partial_\beta   \Psi_\gamma   \Bigr)\Delta \mathcal{X}_\beta + \frac{\partial \mathcal{L}}{\partial (\partial_\alpha  \Psi_\gamma )}\Delta \Psi_\gamma  \biggr] = 0 \,.
\end{aligned}
\end{align}
Here, the expression in the square brackets are the components of the Noether current $I_\alpha $. The term $\partial /\partial \mathcal{X}_\alpha $ in front of $I_\alpha $ should
be interpreted as an implicit derivative with respect to time and three spatial coordinates. The symmetry operations $\Delta \mathcal{X}_\alpha  $ and
$\Delta \Psi_\alpha  $ will be specified below. With
regard to the Lagrangian in Eq.~\eqref{Lag2} we introduce the components $\Psi_\alpha  = ( m_x,m_y,m_z,\sigma_x,\sigma_y,\sigma_z )$ and their partial
derivatives with respect to the independent variables $\partial_\beta\Psi_\alpha  =\partial \Psi_\alpha  /\partial\mathcal{X}_\beta  $. Since we examine an
Euclidean field theory a distinction between upper and lower indices is not necessary. Eq.~\eqref{NT1} can be rewritten by using Eq.~\eqref{variation}. This yields
\begin{align}\label{NT2}
\begin{aligned}
\frac{\partial}{\partial t}\mathcal{L}\Delta t + \frac{\partial}{\partial x_\alpha }\mathcal{L}\Delta x_\alpha +\frac{\partial \mathcal{L}}{\partial \Psi_\alpha }\bigl( \Delta \Psi_\alpha  - \frac{\partial}{\partial t} \Psi_\alpha  \Delta t - \frac{\partial}{\partial x_\beta }
\Psi_\alpha  \Delta x_\beta  \bigr) = 0 \,.
\end{aligned}
\end{align}
In this equation we distinguish between the time and space variables $t$ and $x_\alpha $ explicitly. Eq.~\eqref{NT2} is the basis for the application of the following
symmetry operations. Now we study the rotation around a certain axis as a relevant one. Here we select for instance the $z$-axis. Performing a rotation
in coordinate space with the infinitesimal angle $\Delta \Theta$ the change of the $x$ and $y$-coordinates obeys
\begin{align}
\Delta t = 0 \,,\, \Delta x_\alpha  = \Delta {R}_{\alpha \beta }x_\beta  \, , \, \Delta {R}= \begin{pmatrix} 0 & \Delta \Theta \\ -\Delta\Theta & 0 \end{pmatrix} \,.
\end{align}
In the same manner one can perform the rotation in the configuration space of the moments $\mathbf{m}$ and $\boldsymbol{\sigma}$ symbolized by the before introduced
vector $\Psi_\alpha  = \{ m_x,m_y,m_z,\sigma_x,\sigma_y,\sigma_z\}$. The transformation reads
$\Delta \Psi_\alpha = \Delta {S}_{\alpha \beta}(\Delta \Phi)\Psi _\beta $, where the
rotation matrix is a $6 \times 6$-matrix determined by the rotation angle $\Delta \Phi$. Because both rotations in coordinate space and configuration space, respectively,
are in general independent from each other we find two conserved quantities. Using Eq.~\eqref{NT2} it results
\begin{align}\label{ConRot}
\hat{D}_z\mathcal{L} = 0\,,\quad \hat{\Gamma}_z \mathcal{L} =0\,.
\end{align}
Here the two operators $\hat{D}_z$ and $\hat{\Gamma}_z$ are expressed by
\begin{align}\label{ConRotOp1}
\begin{aligned}
\hat{D}_z &= \hat{L}_z  - \left( \hat{L}_z\psi_\alpha \right) \frac{\partial}{\partial \Psi_\alpha } \,, \\
\hat{\Gamma}_z &= \hat{S}^{(m)}_z + \hat{S}^{(\sigma)}_z\,.
\end{aligned}
\end{align}
The quantity $\hat{L}_z$ is the generator of an infinitesimal rotation around the $z$-axis in the coordinate space
\begin{align}\label{dreh}
\hat{L}_z = y\frac{\partial}{\partial x} - x\frac{\partial}{\partial y}\,,
\end{align}
and therefore, it is identical with the angular momentum operator.
The other quantities $ \hat{S}^{(m)}_z $ and $\hat{S}^{(\sigma)}_z $ are the corresponding
generators in the configuration space of the moments. They are defined as
\begin{align}\label{spin}
\begin{aligned}
\hat{S}^{(m)}_z &=  m_y\frac{\partial }{\partial m_x }  -  m_x\frac{\partial }{\partial m_y } \,, \\
\hat{S}^{(\sigma)}_z &=  \sigma_y \frac{\partial }{\partial \sigma_x }  -  \sigma_x \frac{\partial }{\partial \sigma_y }  \,.
\end{aligned}
\end{align}
These operators reflect the invariance of the total magnetic moment $\mathbf{m} +  \boldsymbol{\sigma} $ under rotation.
Moreover the system is invariant under the combined transformation expressed by $\hat{D}_z $ and $\hat{\Gamma}_z$, where $\hat{D}_z $
offers due to the coupling between system and bath variables as well as the breaking of time reversal invariance a coupling between magnetic moments and
the angular momentum.


\section{Conclusion}

\noindent In this paper we have presented an approach for a mesoscopic magnetic system with dissipation. The Lagrangian consists of two interacting subsystems
characterized by the active magnetic moments of the system $\mathbf{m}$ and the dynamically inactive moments of the bath denoted as $\boldsymbol{\sigma}$.
Both systems are in contact so that a small local alteration of the system variables $\mathbf{m}$ is related as well to a small change of the bath
variables $\boldsymbol{\sigma}$ and vice versa. Due to this constraint we are able to describe the system by a common Lagrangian which incorporates
both degrees of freedom and their coupling. In case the bath variables are constant then the coupling between both systems is absent and the whole system
decays into two independent subsystems. The magnetic moments $\mathbf{m}$ perform a precession around an effective field which is self-organized by the mutual
interaction of the moments. If the coupling between both subsystems breaks the time reversal symmetry the related evolution equation of the moments $\mathbf{m}$
is associated with the Landau-Lifshitz-Gilbert equation which describes both the precession of magnetic moments as well as their damping. It turned out that the
bath variable $\boldsymbol{\sigma}$ can be linked to the effective magnetic field which drives the motion of the
magnetic moments. As consequence the motion of the moments is influenced by the additional bath degrees of freedom. This influence is formulated mathematically and
is described by an inhomogeneous diffusion equation. Finally, we have found conservation laws by means of symmetry considerations based on Noether's theorem. Aside
from the expected symmetry transformation in the coordinate space and the configuration space of the moments, the analysis offers
in a non-relativistic Euclidean field theory an unexpected coupling between both. This point deserves further consideration.
Our approach could be also considered as starting point for a further analysis in magnetic and multiferroic systems. Especially, we are interested in more refined
models which include for instance higher order couplings or anisotropy in the Lagrangian. In multiferroic systems one could study the case that the magnetic and the
polar subsystem have their own reservoirs.\\

One of us (T.B.) is grateful to the Research Network 'Nanostructured Materials'\,,
which is supported by the Saxony-Anhalt State, Germany.

\clearpage


\bibliography{LitNoeth}

\begin{thebibliography}{20}%
\makeatletter
\providecommand \@ifxundefined [1]{%
 \@ifx{#1\undefined}
}%
\providecommand \@ifnum [1]{%
 \ifnum #1\expandafter \@firstoftwo
 \else \expandafter \@secondoftwo
 \fi
}%
\providecommand \@ifx [1]{%
 \ifx #1\expandafter \@firstoftwo
 \else \expandafter \@secondoftwo
 \fi
}%
\providecommand \natexlab [1]{#1}%
\providecommand \enquote  [1]{``#1''}%
\providecommand \bibnamefont  [1]{#1}%
\providecommand \bibfnamefont [1]{#1}%
\providecommand \citenamefont [1]{#1}%
\providecommand \href@noop [0]{\@secondoftwo}%
\providecommand \href [0]{\begingroup \@sanitize@url \@href}%
\providecommand \@href[1]{\@@startlink{#1}\@@href}%
\providecommand \@@href[1]{\endgroup#1\@@endlink}%
\providecommand \@sanitize@url [0]{\catcode `\\12\catcode `\$12\catcode
  `\&12\catcode `\#12\catcode `\^12\catcode `\_12\catcode `\%12\relax}%
\providecommand \@@startlink[1]{}%
\providecommand \@@endlink[0]{}%
\providecommand \url  [0]{\begingroup\@sanitize@url \@url }%
\providecommand \@url [1]{\endgroup\@href {#1}{\urlprefix }}%
\providecommand \urlprefix  [0]{URL }%
\providecommand \Eprint [0]{\href }%
\providecommand \doibase [0]{http://dx.doi.org/}%
\providecommand \selectlanguage [0]{\@gobble}%
\providecommand \bibinfo  [0]{\@secondoftwo}%
\providecommand \bibfield  [0]{\@secondoftwo}%
\providecommand \translation [1]{[#1]}%
\providecommand \BibitemOpen [0]{}%
\providecommand \bibitemStop [0]{}%
\providecommand \bibitemNoStop [0]{.\EOS\space}%
\providecommand \EOS [0]{\spacefactor3000\relax}%
\providecommand \BibitemShut  [1]{\csname bibitem#1\endcsname}%
\let\auto@bib@innerbib\@empty
\bibitem [{\citenamefont {Feynman}\ \emph {et~al.}(2010)\citenamefont
  {Feynman}, \citenamefont {Hibbs},\ and\ \citenamefont
  {Styer}}]{feynman:2010:Book:QMandPI}%
  \BibitemOpen
  \bibfield  {author} {\bibinfo {author} {\bibfnamefont {R.}~\bibnamefont
  {Feynman}}, \bibinfo {author} {\bibfnamefont {A.}~\bibnamefont {Hibbs}}, \
  and\ \bibinfo {author} {\bibfnamefont {D.}~\bibnamefont {Styer}},\
  }\href@noop {} {\emph {\bibinfo {title} {Quantum Mechanics and Path
  Integrals: Emended Edition}}}\ (\bibinfo  {publisher} {Dover Publications},\
  \bibinfo {year} {2010})\BibitemShut {NoStop}%
\bibitem [{\citenamefont {Kleinert}(2009)}]{kleinert:2009:book:pathintegrals}%
  \BibitemOpen
  \bibfield  {author} {\bibinfo {author} {\bibfnamefont {H.}~\bibnamefont
  {Kleinert}},\ }\href@noop {} {\emph {\bibinfo {title} {Path integrals in
  quantum mechanics, statistics, polymer physics, and financial markets}}}\
  (\bibinfo  {publisher} {World Scientific},\ \bibinfo {year}
  {2009})\BibitemShut {NoStop}%
\bibitem [{\citenamefont {Weiss}(1999)}]{weiss:1999:book:quantdissip}%
  \BibitemOpen
  \bibfield  {author} {\bibinfo {author} {\bibfnamefont {U.}~\bibnamefont
  {Weiss}},\ }\href@noop {} {\emph {\bibinfo {title} {{Quantum dissipative
  systems}}}},\ Series in modern condensed matter physics\ (\bibinfo
  {publisher} {World Scientific},\ \bibinfo {year} {1999})\BibitemShut
  {NoStop}%
\bibitem [{\citenamefont {Schulman}(1968)}]{Schulman:PhysRev176:1558:1968}%
  \BibitemOpen
  \bibfield  {author} {\bibinfo {author} {\bibfnamefont {L.}~\bibnamefont
  {Schulman}},\ }\href {\doibase 10.1103/PhysRev.176.1558} {\bibfield
  {journal} {\bibinfo  {journal} {Phys. Rev.}\ }\textbf {\bibinfo {volume}
  {176}},\ \bibinfo {pages} {1558} (\bibinfo {year} {1968})}\BibitemShut
  {NoStop}%
\bibitem [{\citenamefont {Cabra}\ \emph {et~al.}(1997)\citenamefont {Cabra},
  \citenamefont {Dobry}, \citenamefont {Greco},\ and\ \citenamefont
  {Rossini}}]{Cabra:JPA30:2699:1997}%
  \BibitemOpen
  \bibfield  {author} {\bibinfo {author} {\bibfnamefont {D.~C.}\ \bibnamefont
  {Cabra}}, \bibinfo {author} {\bibfnamefont {A.}~\bibnamefont {Dobry}},
  \bibinfo {author} {\bibfnamefont {A.}~\bibnamefont {Greco}}, \ and\ \bibinfo
  {author} {\bibfnamefont {G.~L.}\ \bibnamefont {Rossini}},\ }\href@noop {}
  {\bibfield  {journal} {\bibinfo  {journal} {J. Phys. A}\ }\textbf {\bibinfo
  {volume} {30}},\ \bibinfo {pages} {2699} (\bibinfo {year}
  {1997})}\BibitemShut {NoStop}%
\bibitem [{\citenamefont {Smirnov}(1999)}]{Smirnov:JPA32:1285:1999}%
  \BibitemOpen
  \bibfield  {author} {\bibinfo {author} {\bibfnamefont {V.~V.}\ \bibnamefont
  {Smirnov}},\ }\href@noop {} {\bibfield  {journal} {\bibinfo  {journal} {J.
  Phys. A}\ }\textbf {\bibinfo {volume} {32}},\ \bibinfo {pages} {1285}
  (\bibinfo {year} {1999})}\BibitemShut {NoStop}%
\bibitem [{\citenamefont {Grinberg}(2003)}]{Grinberg:PLA311:133:2003}%
  \BibitemOpen
  \bibfield  {author} {\bibinfo {author} {\bibfnamefont {H.}~\bibnamefont
  {Grinberg}},\ }\href {\doibase 10.1016/S0375-9601(03)00464-X} {\bibfield
  {journal} {\bibinfo  {journal} {Phys. Lett. A}\ }\textbf {\bibinfo {volume}
  {311}},\ \bibinfo {pages} {133} (\bibinfo {year} {2003})}\BibitemShut
  {NoStop}%
\bibitem [{\citenamefont {Koch}\ \emph {et~al.}(2008)\citenamefont {Koch},
  \citenamefont {Gro\ss{}mann}, \citenamefont {Stockburger},\ and\
  \citenamefont {Ankerhold}}]{Koch:PhysRevLett.100.230402:2008}%
  \BibitemOpen
  \bibfield  {author} {\bibinfo {author} {\bibfnamefont {W.}~\bibnamefont
  {Koch}}, \bibinfo {author} {\bibfnamefont {F.}~\bibnamefont {Gro\ss{}mann}},
  \bibinfo {author} {\bibfnamefont {J.~T.}\ \bibnamefont {Stockburger}}, \ and\
  \bibinfo {author} {\bibfnamefont {J.}~\bibnamefont {Ankerhold}},\ }\href
  {\doibase 10.1103/PhysRevLett.100.230402} {\bibfield  {journal} {\bibinfo
  {journal} {Phys. Rev. Lett.}\ }\textbf {\bibinfo {volume} {100}},\ \bibinfo
  {pages} {230402} (\bibinfo {year} {2008})}\BibitemShut {NoStop}%
\bibitem [{\citenamefont {Landau}\ and\ \citenamefont
  {Lifshitz}(1935)}]{Landau:ZdS:8:p153:1935}%
  \BibitemOpen
  \bibfield  {author} {\bibinfo {author} {\bibfnamefont {L.}~\bibnamefont
  {Landau}}\ and\ \bibinfo {author} {\bibfnamefont {E.}~\bibnamefont
  {Lifshitz}},\ }\href@noop {} {\bibfield  {journal} {\bibinfo  {journal}
  {Zeitschr. d. Sowj.}\ }\textbf {\bibinfo {volume} {8}},\ \bibinfo {pages}
  {153} (\bibinfo {year} {1935})}\BibitemShut {NoStop}%
\bibitem [{\citenamefont {Gilbert}(2004)}]{Gilbert:ITOM:40:p3443:2004}%
  \BibitemOpen
  \bibfield  {author} {\bibinfo {author} {\bibfnamefont {T.~L.}\ \bibnamefont
  {Gilbert}},\ }\href@noop {} {\bibfield  {journal} {\bibinfo  {journal} {IEEE
  Trans. Magn.}\ }\textbf {\bibinfo {volume} {40}},\ \bibinfo {pages} {3443}
  (\bibinfo {year} {2004})}\BibitemShut {NoStop}%
\bibitem [{\citenamefont {Tserkovnyak}\ \emph {et~al.}(2005)\citenamefont
  {Tserkovnyak}, \citenamefont {Brataas}, \citenamefont {Bauer},\ and\
  \citenamefont {Halperin}}]{Tserkovnyak:RoMP:77:p1375:2005}%
  \BibitemOpen
  \bibfield  {author} {\bibinfo {author} {\bibfnamefont {Y.}~\bibnamefont
  {Tserkovnyak}}, \bibinfo {author} {\bibfnamefont {A.}~\bibnamefont
  {Brataas}}, \bibinfo {author} {\bibfnamefont {G.~E.~W.}\ \bibnamefont
  {Bauer}}, \ and\ \bibinfo {author} {\bibfnamefont {B.~I.}\ \bibnamefont
  {Halperin}},\ }\href@noop {} {\bibfield  {journal} {\bibinfo  {journal} {Rev.
  Mod. Phys.}\ }\textbf {\bibinfo {volume} {77}},\ \bibinfo {pages} {1375}
  (\bibinfo {year} {2005})}\BibitemShut {NoStop}%
\bibitem [{\citenamefont {Durand}\ and\ \citenamefont
  {Paidarová}(2010)}]{Durand:EPL89:67004:2010}%
  \BibitemOpen
  \bibfield  {author} {\bibinfo {author} {\bibfnamefont {P.}~\bibnamefont
  {Durand}}\ and\ \bibinfo {author} {\bibfnamefont {I.}~\bibnamefont
  {Paidarová}},\ }\href@noop {} {\bibfield  {journal} {\bibinfo  {journal}
  {EPL}\ }\textbf {\bibinfo {volume} {89}},\ \bibinfo {pages} {67004} (\bibinfo
  {year} {2010})}\BibitemShut {NoStop}%
\bibitem [{\citenamefont {Safonov}\ and\ \citenamefont
  {Bertram}(2005)}]{Safonov:PhysRevB.71.224402:2005}%
  \BibitemOpen
  \bibfield  {author} {\bibinfo {author} {\bibfnamefont {V.~L.}\ \bibnamefont
  {Safonov}}\ and\ \bibinfo {author} {\bibfnamefont {H.~N.}\ \bibnamefont
  {Bertram}},\ }\href {\doibase 10.1103/PhysRevB.71.224402} {\bibfield
  {journal} {\bibinfo  {journal} {Phys. Rev. B}\ }\textbf {\bibinfo {volume}
  {71}},\ \bibinfo {pages} {224402} (\bibinfo {year} {2005})}\BibitemShut
  {NoStop}%
\bibitem [{\citenamefont {Weetman}\ and\ \citenamefont
  {Akhras}(2009)}]{Weetman:JAP105:2009}%
  \BibitemOpen
  \bibfield  {author} {\bibinfo {author} {\bibfnamefont {P.}~\bibnamefont
  {Weetman}}\ and\ \bibinfo {author} {\bibfnamefont {G.}~\bibnamefont
  {Akhras}},\ }\href {\doibase 10.1063/1.3073896} {\bibfield  {journal}
  {\bibinfo  {journal} {J. Appl. Phys.}\ }\textbf {\bibinfo {volume} {105}},\
  \bibinfo {pages} {023917} (\bibinfo {year} {2009})}\BibitemShut {NoStop}%
\bibitem [{\citenamefont {Gareeva}\ and\ \citenamefont
  {Zvezdin}(2010)}]{Gareeva:EPL91:47006:2010}%
  \BibitemOpen
  \bibfield  {author} {\bibinfo {author} {\bibfnamefont {Z.~V.}\ \bibnamefont
  {Gareeva}}\ and\ \bibinfo {author} {\bibfnamefont {A.~K.}\ \bibnamefont
  {Zvezdin}},\ }\href@noop {} {\bibfield  {journal} {\bibinfo  {journal} {EPL}\
  }\textbf {\bibinfo {volume} {91}},\ \bibinfo {pages} {47006} (\bibinfo {year}
  {2010})}\BibitemShut {NoStop}%
\bibitem [{\citenamefont {Ovchinnikov}\ and\ \citenamefont
  {Wang}(2010)}]{Ovchinnikov:PhysRevB.82.024410}%
  \BibitemOpen
  \bibfield  {author} {\bibinfo {author} {\bibfnamefont {I.~V.}\ \bibnamefont
  {Ovchinnikov}}\ and\ \bibinfo {author} {\bibfnamefont {K.~L.}\ \bibnamefont
  {Wang}},\ }\href {\doibase 10.1103/PhysRevB.82.024410} {\bibfield  {journal}
  {\bibinfo  {journal} {Phys. Rev. B}\ }\textbf {\bibinfo {volume} {82}},\
  \bibinfo {pages} {024410} (\bibinfo {year} {2010})}\BibitemShut {NoStop}%
\bibitem [{\citenamefont
  {Jackson}(1999)}]{jackson:1999:book:classicalelectrodynamics}%
  \BibitemOpen
  \bibfield  {author} {\bibinfo {author} {\bibfnamefont {J.}~\bibnamefont
  {Jackson}},\ }\href@noop {} {\emph {\bibinfo {title} {Classical
  electrodynamics}}}\ (\bibinfo  {publisher} {Wiley},\ \bibinfo {year}
  {1999})\BibitemShut {NoStop}%
\bibitem [{\citenamefont {Landau}\ \emph {et~al.}(1980)\citenamefont {Landau},
  \citenamefont {Lifshitz},\ and\ \citenamefont
  {Pitaevskii}}]{Landau:StatPhysPart2:Book:1980}%
  \BibitemOpen
  \bibfield  {author} {\bibinfo {author} {\bibfnamefont {L.~D.}\ \bibnamefont
  {Landau}}, \bibinfo {author} {\bibfnamefont {E.}~\bibnamefont {Lifshitz}}, \
  and\ \bibinfo {author} {\bibfnamefont {L.}~\bibnamefont {Pitaevskii}},\
  }\href@noop {} {\emph {\bibinfo {title} {Statistical Physics Part 2: Theory
  of the Condensed State}}}\ (\bibinfo  {publisher} {Pergamon Press},\ \bibinfo
  {address} {Oxford},\ \bibinfo {year} {1980})\BibitemShut {NoStop}%
\bibitem [{\citenamefont {Noether}(1918)}]{Noether:Ges.Wiss.Goettingen:1918}%
  \BibitemOpen
  \bibfield  {author} {\bibinfo {author} {\bibfnamefont {E.}~\bibnamefont
  {Noether}},\ }\href@noop {} {\bibfield  {journal} {\bibinfo  {journal}
  {Nachr. Ges. Wiss. G\"ottingen}\ ,\ \bibinfo {pages} {235}} (\bibinfo {year}
  {1918})}\BibitemShut {NoStop}%
\bibitem [{\citenamefont {Hill}(1951)}]{Hill:ConservationTheorems:1951}%
  \BibitemOpen
  \bibfield  {author} {\bibinfo {author} {\bibfnamefont {E.~L.}\ \bibnamefont
  {Hill}},\ }\href {\doibase 10.1103/RevModPhys.23.253} {\bibfield  {journal}
  {\bibinfo  {journal} {Rev. Mod. Phys.}\ }\textbf {\bibinfo {volume} {23}},\
  \bibinfo {pages} {253} (\bibinfo {year} {1951})}\BibitemShut {NoStop}%
\end{thebibliography}%
\bibliographystyle{apsrev4-1}

\end{document}